\documentclass[11pt]{article}
\usepackage{amsmath,amssymb}
\usepackage{graphicx}
\usepackage{cancel}
\usepackage{epsfig}
\usepackage{float}
\usepackage{subfigure}
\usepackage[hang]{footmisc}
\begin{document}
\oddsidemargin .03in
\evensidemargin 0 true pt
\topmargin -.4in
 
\def\ra{{\rightarrow}} 
\def\a{{\alpha}}
\def\g{{\gamma}}
\def\o{{\omega}}
\def\s{{\sigma}}
\def\b{{\beta}}
\def\l{{\lambda}}
\def\eps{{\epsilon}}
\def\tri{{\triangle}}
\def\pr{{\partial}}
\def\na{{\nabla }}
\def\sp{\vspace{.15in}}
\def\hs{\hspace{.25in}}
\def\n{\nonumber}
\def\ni{{\noindent}}
\def\Ra{{\Rightarrow}}
\def\be{\begin{equation}}
\def\ee{\end{equation}}
\def\bea{\begin{eqnarray}}
\def\eea{\end{eqnarray}}


\begin{titlepage}
\topmargin= -.2in
\textheight 9.5in

\begin{center}
\baselineskip= 18 truept

\vspace{.3in}

\centerline{\Large\bf A review of Einstein Cartan Theory}
\centerline{\Large\bf to describe superstrings with intrinsic torsion }

\vspace{.6in}
\noindent
{\bf Richa Kapoor}\footnote{richa.phy@gmail.com }

\vspace{.2in}

\noindent

\noindent
{{\Large Department of Physics \& Astrophysics}\\
{\Large University of Delhi, New Delhi 110 007, India}}

\vspace{.2in}

{\today}
\thispagestyle{empty}

\vspace{.6in}
\begin{abstract}

\baselineskip=14 truept  

\vspace{.12in}
This paper reviews the Einstein Cartan theory (ECT), the famous extension of general relativity (GR) in presence of spacetime torsion. The vacuum equations are derived step by step. Vielbein formulation is discussed for determining the field equations in presence of matter. This review would be easily comprehensible for any student familiar with general relativity. Further, ECT is used to describe superstrings with intrinsic torsion, assuming a $D_p$-brane in presence of a curved background of the NS-NS Kalb-Ramond field. D-brane worldvolume is a flat spacetime governed by the Dirac-Born-Infeld (DBI) action. In presence of the dynamical NS-NS $B$-field, the contortion tensor equals the totally antisymmetric torsion. Using this, the form of the $D_p$-brane action in presence of torsion is determined.

\vspace{1in}

\noindent

\noindent

\end{abstract}
\end{center}

\vspace{.2in}

\baselineskip= 16 truept

\vspace{1in}

\end{titlepage}

\baselineskip= 18 truept
   
\section{Motivation}
Einstein Cartan Theory (ECT) is an extension of General Relativity (GR) theory, which is  the simplest theory of gravity with curvature as the only geometric property of the spacetime. General relativity is a classical theory designed by Einstein on a pseudo-Riemannian\footnote{a pseudo-Riemannian manifold is a generalization of a Riemannian manifold in which the metric tensor need not be positive-definite} manifold. On the other hand, ECT has  curvature and torsion both as the geometric properties of the spacetime. Motivation to devise this extension arose by comparing general relativity with theories of the other three fundamental interactions. Strong, weak and electromagnetic forces are described by quantum relativistic fields  in a flat Minkowski space. The spacetime itself is unaffected by these fields. On the contrary, gravitational interactions  modify the geometrical structure of spacetime and they are not represented by another field but by the distortion of  geometry itself \cite{sabb}. While three-fourth of modern physics acting at a microscopic level is described in the framework of flat spacetime, the remaining one-fourth i.e. the macroscopic physics of  gravity needs introduction of a dynamic or geometrical background. This situation is inadequate because three fundamental interactions are completely disjoint from the remaining one. So a theory needs to be formulated, which can in some limit  give common description for all the four. In other words, the problem is {\it what if we consider elementary particle interactions in a curved  spacetime}. A big drawback of general relativity is that it assumes matter to be mass energy distribution but actually matter also includes spin density.

\sp
\ni
For macroscopic objects, spin averages out in general if we ignore objects like ferromagnets but at microscopic level, spin plays an important role. Since gravity is the weakest interaction at low energy, it appears that gravitation has no effect on the elementary particle interactions. However, when we consider microphysics in curved spacetime, we have some important phenomena like neutron interferometry which can be used to observe the interaction of neutrons with Earth's gravitational field \cite{colella}. Macroscopically, spin density plays significant role in early universe (big bang) and	superdense objects like neutron stars and black holes.

\sp
\ni
A mass distribution in a spacetime is described by the energy-momentum tensor while a spin distribution in a field theory is described by 
the spin density tensor. So at the microscopic level, energy-momentum tensor is not sufficient to characterize the matter sources but the spin density tensor is also needed. However, if we consider a system of scalar fields depicting spinless particles, the spin density tensor vanishes. 

\sp
\ni 
Similar to the mass-energy distribution (a property of matter) which produces curvature in spacetime (a geometric property), spin density must also couple to some  geometric property.  That property should be torsion. ECT is GR extended to include torsion. It is also called ECSK theory (after Einstein, Cartan, 
Sciama, Kibble who laid the foundations of this theory), and briefly denoted as $U_4$ theory, where $U_4$ is a four dimensional Riemann-Cartan spacetime.
 Torsion leads to deviation from general relativity only in exceptional situations like big bang, gravitational collapse and microscopic physics.
 
\section{Spacetime torsion}
In GR, we interpret gravity not as a force but as the curvature or bending of spacetime produced by a mass energy distribution. We have the constraint of torsion free spacetime, hence the  connection is symmetric. In ECT, it is assumed that in addition we have a spin density of matter which produces torsion in spacetime around and connection is in general asymmetric. Then torsion is the antisymmetric part of the connection

\be  {Q_{\mu\nu}}^{\a} = \frac{1}{2}(\Gamma^{\a}_{\mu\nu}-\Gamma^{\a}_{\nu\mu})=\Gamma^{\a}_{[\mu\nu]}\ee

\ni
Torsion ${Q_{\mu\nu}}^{\a}$  is a third-rank tensor with antisymmetry in its first two indices. It has ${D^2}(D-1)/2$ independent components in a $D$ dimensional spacetime. 

\sp\ni
{\bf Effect of torsion on the geometry of spacetime}
\begin{figure}[H]
 \centering{\includegraphics[width=.76\textwidth,height=.48\textheight]{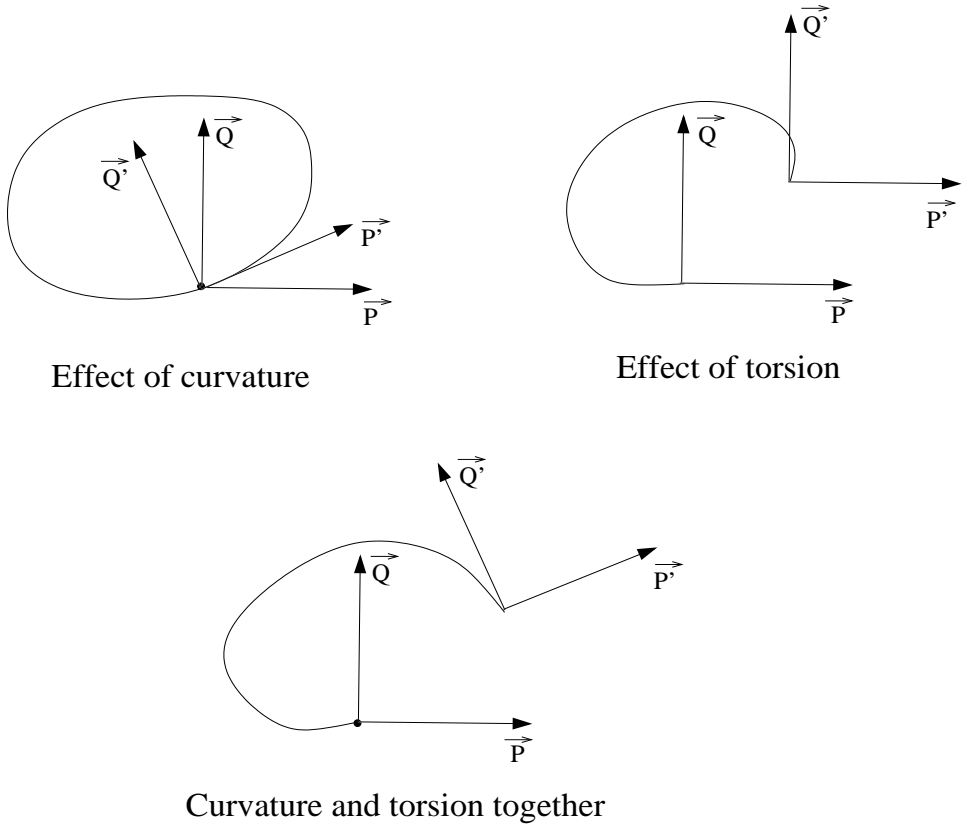}}
\caption{\small{\bf\it When an infinitesimal or tangent vector is parallel transported along a closed path, we have a rotation (if there is only curvature) or a translation (if there is only torsion) or both (if there is curvature and torsion)} \cite{sabb} }\label{curvator}
\end{figure} 

\ni
To understand the geometrical meaning of torsion, we compare it with the intrinsic curvature of GR. When a tangent vector is parallel transported along a closed path, it changes its direction. But, in presence of torsion, if we try to parallel transport it along a closed path, it would come back translated with respect to its original position i.e. path will not be closed. This is illustrated in the figure \ref{curvator}.

\section{Metric Compatibility}
In general relativity, there are two constraints: (1.) metric compatibility of the affine connection and (2.) torsion free spacetime, and hence the connection is symmetric, i.e., Christoffel connection.
If we relax both these constraints then what we have is a general affine manifold, $A_4$. For $A_4$, the affine connection is
\be \left.{\Gamma}^{{\alpha}{(A_4)}}_{\mu\nu} = \left\{^\a _{\mu\nu}\right\} - {K_{\mu\nu}}^\alpha - {V_{\mu\nu}}^\alpha \right. \label{a4}\ee
where $\left\{^\a _{\mu\nu}\right\}$ is called the Christoffel symbol, ${K_{\mu\nu}}^\alpha$ is known as contortion tensor and ${V_{\mu\nu}}^\alpha$ arises from the non-metricity.

\sp\ni {\bf Contortion tensor:}  Torsion appears in linear combination as the contortion tensor ${K_{\mu\nu}}^\alpha$ \be {K_{\mu\nu}}^\alpha = - {Q_{\mu\nu}}^\alpha + {Q_\nu}{^\alpha}_\mu - {Q^\alpha}_{\mu\nu} \Rightarrow {K_{\mu\nu}}^\alpha = - {K_\mu}{^\alpha}_\nu \label{contort}\ee It is antisymmetric in 2nd and 3rd indices.  Another important combination is the {\it modified torsion tensor}
\be { T_{\mu\nu}}^\rho = {Q_{\mu\nu}}^\rho + 2 \delta^\rho_{[\mu} Q_{\nu]}\ee

\sp\ni {\bf Non-metricity:} In eq. (\ref{a4}),
\be V_{\mu\nu\alpha} = \dfrac{1}{2} \left(D_\alpha g_{\mu\nu} - D_\nu g_{\mu\alpha} - D_\mu g_{\nu\alpha}\right)\ee  with $D^{(A_4)}_\alpha g_{\mu\nu}\neq 0$ known as the {\bf non-metricity tensor} and the covariant derivative of the affine manifold is defined by \be D_\a^{(A_4)} = \partial_\a +  \Gamma_\a^{(A_4)} \ .\ee
 However in a Riemann-Cartan or $U_4$ manifold, one constraint is relaxed, that of the torsion free spacetime. Metric compatibility condition is still there in $U_4$, i.e.,
\bea  D_\alpha g_{\mu\nu} = 0  \Rightarrow V_{\mu\nu\alpha}=0\ \textrm{ hence }\ D_\a^{(U_4)} &=& \partial_\a + \Gamma_\a^{(U_4)}  \n\\ \textrm{ with } \ {\Gamma}^{{\alpha}{(U_4)}}_{\mu\nu} &=&\left\{^\a _{\mu\nu}\right\} - {K_{\mu\nu}}^\alpha  \ . \eea 
As a result of metric compatibility, the unit angles and lengths are preserved. Metric is covariantly constant so the lengths of the measuring rods and the angles between two of them do not change under parallel transfer. It saves a locally Minkowskian structure of the spacetime. Since the Riemann-Cartan manifold is unit preserving, so it is also called $U_4$ manifold.

\sp
\ni
  Metric compatibility condition in $U_4$ also implies metric compatibility in $V_4$, i.e. the Riemann manifold as follows
\bea
&&D^{(U_4)}_\alpha g^{\mu\nu} = 0\n\\
&\Rightarrow & \partial_\alpha g^{\mu\nu} + {\Gamma} ^\mu_{\alpha\rho}   g^{\rho\nu} + {\Gamma} ^\nu_{\alpha\rho}  g^{\mu\rho} = 0 \n\\
&\Rightarrow & \partial_\alpha g^{\mu\nu} + \left\{^\mu _{\a\rho}\right\} g^{\rho\nu} -\ \nearrow \hspace{-0.24in} {K _{\alpha\rho}}^\mu g^{\rho\nu} + \left\{^\nu _{\a\rho}\right\} g^{\rho\mu} -\ \nearrow \hspace{-0.24in} {K _{\alpha\rho}}^\nu g^{\rho\mu}  = 0 \n\\
&\Rightarrow & \nabla _\alpha ^{(V_4)} g^{\mu\nu}
 = 0\ .\eea
 
\sp \ni
 {\bf Trace-free contortion tensor}\\
Trace of contortion tensor ${K_{\mu\nu}}^\alpha = - {Q_{\mu\nu}}^\alpha + {Q_\nu}{^\alpha}_\mu - {Q^\alpha}_{\mu\nu} $ 
 over its various indices gives
 \be {{K^\alpha}_\nu}^\nu = 0,\ {{K^\alpha}_\alpha}^\rho = 2 Q^\rho \textrm{ and } {K^{\alpha\rho}}_\alpha = -2 Q^\rho\ee
$Q^\rho$ is the torsion vector. Traceless part of the contortion tensor is \bea &\tilde K_{\mu\nu\alpha} &= K_{\mu\nu\alpha} + \dfrac{2}{3} (g_{\mu\alpha} Q_\nu - g_{\mu\nu} Q_\alpha), \n\\ 
\textrm{since its trace is }  &g^{\mu\nu}\tilde K_{\mu\nu\alpha}& = {\tilde K^\mu}_{\ \;\mu\alpha} = 2 Q_\alpha + \dfrac{2}{3} (Q_\alpha - 4 Q_\alpha) = 0 . \label{con1}\eea
 
 \section{Autoparallels and extremals}
 When we study the curves of choice in a Riemann-Cartan spacetime, we must distinguish between the two classes of curves both of which reduce to the geodesics of the Riemannian space when we set torsion equal to zero \cite{hhkn}.
 
 \sp
 \ni
{\bf Autoparallel curves }(straightest lines) are curves over which a vector is transported parallel to itself, according to the affine connection of the manifold. Parallel displacement of a vector $A^\mu$ from                                                                                                                                                                                                                                                                                                                                                                                                                                                                                                                                                                                                                                                                                                                                  $x^\rho$ to $x^\rho + dx^\rho$ leads to \be dA^\mu = - {\Gamma}^\mu_{\nu\rho} A^\nu dx^\rho \ee
 Using this equation with a chosen suitable affine parameter $s$, we get the differential equation of the autoparallels 
\be  \frac{d^2x^\alpha}{ds^2} + {\Gamma} ^\alpha_{(\mu\nu)} \frac{dx^\mu}{ds} \frac{dx^\nu}{ds} = 0 \ee 
where $ {\Gamma} ^\alpha_{(\mu\nu)} =  \left\{^\a _{\mu\nu}\right\} - {K_{(\mu\nu)}}^\alpha = \left\{^\a _{\mu\nu}\right\} + 2 {Q^\alpha_{(\mu\nu)}} $. \\
Notice that only the symmetric (but torsion dependent) part of the connection enters in this equation, because of symmetry of the product $dx^\mu dx^\nu = dx^\nu dx^\mu $.

\sp
\ni
{\bf Extremal curves} (shortest or longest lines) are curves which are of extremal length with respect to the metric of the manifold. According to 
$ds^2 = - g_{\mu\nu} dx^\mu dx^\nu$, length between two points depends only on the metric field (and not on the torsion). Differential equation for the extremals can be derived from \be\delta \int ds = \delta \int\sqrt {- g_{\mu\nu} dx^\mu dx^\nu}  = 0\ee
exactly as in the corresponding Riemannian space and we get 
\be  \frac{d^2x^\alpha}{ds^2} + \left\{^\a _{\mu\nu}\right\} \frac{dx^\mu}{ds} \frac{dx^\nu}{ds} = 0 . \ee 
\ni
In $U_4$, the autoparallels and extremals coincide iff the torsion is totally antisymmetric i.e. $Q_{\mu\nu\rho} = Q_{[\mu\nu\rho]}$.

\section{Parallel or compatible volume element in $U_4$ manifold}
In order to define a general covariant volume element in a manifold, it is necessary to introduce a density quantity $f(x)$ so that $d^4x \rightarrow f(x)d^4x = dvol. $
This is done in order to compensate the Jacobian that arises from the transformation law of the usual volume element $d^4x$ under a coordinate transformation. In GR, the density $f(x) = \sqrt{-g}$ is taken for this purpose. 
In $V_4$, the volume element $\sqrt{-g}\;d^4x$ is said to be compatible with the connection since the scalar density $\sqrt{-g}$ obeys $\nabla_\mu^{(V_4)}\sqrt{-g} = 0 $ where $\nabla_\mu^{(V_4)} = \partial_\mu - \left\{^\a _{\mu\a}\right\} $. 

\sp
\ni
But the same volume element, $\sqrt{-g}\;d^4x$ is not compatible in $U_4$ since
\[D^{(U_4)}_\mu \sqrt{-g} = \nabla_\mu^{(V_4)} \sqrt{-g} - 2 Q_\mu \sqrt{-g} = - 2 Q_\mu \sqrt{-g} \neq 0\ .\]
In order to define such \textit{parallel} volume element in $U_4$ manifolds, one needs to find out a covariantly constant density $f(x)$. Such density exists only if the torsion vector, $Q_\mu$, can be obtained from a scalar potential $Q_\mu (x) = \partial_\mu \Theta(x)$. In this case we have \[ f(x) = e^{2 \Theta}\sqrt{-g} \] 
\bea \Rightarrow  D^{(U_4)}_\mu f(x)&=& \partial _\mu f(x) - {\Gamma} ^\rho _{\rho\mu} f(x) \n\\
  &=& \partial _\mu (e^{2 \Theta}\sqrt{-g}) - \left\{^\a _{\mu\a}\right\} e^{2 \Theta}\sqrt{-g} +  {K_{\rho\mu}}^\rho e^{2 \Theta}\sqrt{-g}\n\\
   &=& 0\eea
   So, $dvol = e^{2 \Theta}\sqrt{-g}\; d^4x$ is the volume element compatible with the connection in Riemann Cartan manifolds.
   
   \sp\ni 
   {\bf Generalised Gauss's law in $U_4$ } 
  \bea \int dvol\; D_\mu A^\mu &=& \int d^4x\; e^{2 \Theta}\sqrt{-g}\; D_\mu A^\mu\n\\ &=& \int d^4x\; \partial_\mu (e^{2 \Theta}\sqrt{-g} A^\mu) = \textrm{surface term} \eea
 where ${\Gamma} ^\rho _{\rho\mu} = \partial_\mu \; ln \; (e^{2 \Theta}\sqrt{-g}\;) $.
 
 \section{Covariant derivative commutator in $U_4$ manifold}
 In a torsion free space, the covariant derivatives commute in their action on a scalar field. But in the presence of torsion, the commutator  acts on a scalar field $\phi$ as proportional to its first derivative 
\bea [D_\mu, D_\nu] \phi &=& D_\mu\partial_\nu\phi - D_\nu\partial _\mu\phi \n\\  &=&  \partial_\mu\partial_\nu \phi - \Gamma^\rho_{\mu\nu}\partial_\rho\phi - \partial_\nu\partial_\mu \phi + \Gamma^\rho_{\nu\mu}\partial_\rho\phi \n\\ &=& 2 {K_{[\mu\nu]}}^\rho \partial_\rho\phi\ .\eea
 Action of the commutator on a vector field $V^\rho$ is evaluated as follows,
\bea   D_\mu D_\nu V^\rho &=& \partial _\mu (D_\nu V^\rho) - {\Gamma} ^\alpha_{\mu\nu} D_\alpha V^\rho + {\Gamma} ^\rho_{\mu\alpha} D_\nu V^\alpha \n\\
&=& \partial _\mu (\partial_\nu V^\rho + {\Gamma} ^\rho_{\nu\alpha} V^\alpha ) - {\Gamma} ^\alpha_{\mu\nu}(\partial_\alpha V^\rho + {\Gamma} ^\rho_{\alpha\sigma} V^\sigma ) \n\\ &&+\;  {\Gamma} ^\rho_{\mu\alpha} (\partial_\nu V^\alpha + {\Gamma} ^\alpha_{\nu\sigma} V^\sigma )  \n\\  \textrm{Similarly,}\quad -  D_\nu D_\mu V^\rho &=& -\partial _\nu \partial_\mu V^\rho - (\partial _\nu{\Gamma} ^\rho_{\mu\alpha}) V^\alpha  + {\Gamma} ^\alpha_{\nu\mu} \partial_\alpha V^\rho -  {\Gamma} ^\rho_{\mu\alpha} \partial_\nu V^\alpha \n\\ &&\; -  {\Gamma} ^\rho_{\nu\alpha} \partial_\mu V^\alpha +  {\Gamma} ^\alpha_{\nu\mu}{\Gamma} ^\rho_{\alpha\sigma} V^\sigma - {\Gamma} ^\rho_{\nu\alpha} {\Gamma} ^\alpha_{\mu\sigma} V^\sigma   \n\\ 
\Rightarrow [D_\mu, D_\nu] V^\rho &=& {R_{\mu\nu\alpha}}^\rho V^\alpha - 2 {Q_{\mu\nu}}^\alpha D_{\a}V^\rho
\label{rie2}\eea 
\be  \textrm{where }\quad {R_{\mu\nu\alpha}}^\rho \equiv \partial _\mu{\Gamma} ^\rho_{\nu\alpha} - \partial _\nu{\Gamma} ^\rho_{\mu\alpha} + {\Gamma} ^\sigma_{\nu\a} {\Gamma} ^\rho_{\mu\sigma} - {\Gamma} ^\sigma_{\mu\a} {\Gamma} ^\rho_{\nu\sigma}. \label{rie1}\ee
The left hand side of (\ref{rie2}) is manifestly a tensor so ${R_{\mu\nu\alpha}}^\rho $ must be a tensor too, even though it is constructed from non-tensorial segments. ${R_{\mu\nu\alpha}}^\rho $ is the modified curvature tensor in presence of torsion. An important point is that the commutator $[D_\mu, D_\nu]$ has an action on vector fields which is a simple multiplicative transformation in the absence of torsion. The Riemann tensor measures that part of
the commutator of covariant derivatives which is proportional to the vector field, while the torsion tensor measures the part which is proportional to the covariant \textit{first} derivative of the vector field. \textit{Second derivative doesn't occur on the R.H.S.}

\sp\ni
{\bf Symmetry properties of the Riemann curvature tensor in $U_4$}\\
Curvature tensor in $U_4$ has the following antisymmetry properties 
\be R_{\a\mu\nu\s} = - R_{\mu\a\nu\s} = - R_{\a\mu\s\nu} \ee
Antisymmetry between first two indices is easy to see from eqn.(\ref{rie1}), simply with $\mu\leftrightarrow\nu$. To see that between the last two, consider $ R_{\a\b\omega\s} = g_{\rho\s} {R_{\a\b\omega}}^\rho$. After some algebraic manipulations and with $\{\mu,\o\b\} = g_{\mu\nu}  \left\{^\nu _{\o\b}\right\}  $, we get
\bea R_{\a\b\omega\s} &=& R_{\a\b\omega\s}^{(V_4)} - \partial _\a K_{\b\o\s} + \partial _\b K_{\a\o\s} + {K_{\b\o}}^\mu {\{\mu,\s\a}\} - {K_{\a\o}}^\mu {\{\mu,\s\b}\}\n \\  &&+ {K_{\a\s}}^\mu \{{\mu,\o\b}\} - {K_{\b\s}}^\mu {\{\mu,\o\a}\} +  K_{\a\mu\s} {K_{\b\o}}^\mu - {K_{\b\s}}^\mu K_{\a\mu\o}\n\\ &=&  R_{\a\b\omega\s}^{(V_4)} - \nabla _\a K_{\b\o\s} + \nabla_\b K_{\a\o\s} +  K_{\a\mu\s} {K_{\b\o}}^\mu - {K_{\b\s}}^\mu K_{\a\mu\o}\label{rie3}\eea
Thus the curvature can be expressed through the Riemann tensor (of $V_4$) depending only on the metric, covariant derivative $\nabla$ (i.e. torsionless covariant derivative) and contortion tensor.
From this, we easily see $ R_{\a\b\o\s} = - R_{\a\b\s\o} $.

\sp\ni
{\bf Ricci tensor in $U_4$ is asymmetric:}
From (\ref{rie3}),
\bea R_{\b\o} = R{_{\a\b\o}}^\a &=& R_{\b\o}^{(V_4)} - \nabla _\a {K_{\b\o}}^\a + \nabla _\b {K_{\a\o}}^\a + {K_{\a\rho}}^\a   {K_{\b\o}}^\rho  - {K_\b}^{\a\mu} K_{\a\mu\o}\n\\ &=& R_{\b\o}^{(V_4)} - \nabla _\a {K_{\b\o}}^\a - 2 \nabla _\b Q_\o - 2  Q_\rho  {K_{\b\o}}^\rho  - {K_\b}^{\a\mu} K_{\a\mu\o} \label{rie4}
\eea 
Einstein (Cartan) tensor is as usual defined by \be G_{\mu\nu} = R_{\mu\nu} - \dfrac{1}{2} g_{\mu\nu} R .\ee It is also asymmetric in Riemann Cartan space.

\sp\ni
{\bf Ricci scalar in $U_4$} \\
It is useful to work out that \\
a. $\partial _\mu Q^\mu \neq \partial ^\mu Q_\mu$ in curved space. Infact \bea \partial ^\mu Q_\mu = g^{\mu\nu} \partial _\nu Q_\mu = \partial _\mu Q^\mu - Q_\mu \partial _\nu g^{\mu\nu} \eea
b. With $\tilde K _{\a\nu\rho}$ as the tracefree contortion tensor defined in eqn.(\ref{con1}), \be \tilde K_{\nu\rho\a} \tilde K^{\a\nu\rho} = K_{\nu\rho\a} K^{\a\nu\rho}  + \dfrac{4}{3} \partial _\rho \Theta \partial ^\rho \Theta \label{sca1} \ee
So the curvature scalar, from (\ref{rie4}), is 
\bea R = g^{\mu\nu} R_{\mu\nu} &=& R^{(V_4)} + 2 \nabla _\a {K_\mu}^{\a\mu}  + {K_{\a\nu}}^\a {K_\rho}^{\rho\nu} - {K_\rho}^{\a\mu} K{_{\a\mu}}^\rho\n\\ &=& R^{(V_4)}  - 4\nabla_\nu Q^\nu  - 4 Q_\rho Q^\rho - K_{\rho\a\mu} K^{\a\mu\rho} \label{sca2}
\eea
and in terms of tracefree contortion tensor, using eqns. (\ref{sca1}) and (\ref{sca2})
\bea \Rightarrow R = R^{(V_4)} - 4 D_\nu \partial^\nu\Theta + \dfrac{16}{3}\; \partial_\rho\Theta\; \partial^\rho \Theta - \tilde K_{\nu\rho\a} \tilde K^{\a\nu\rho} \eea

\sp\ni
{\bf Einstein Hilbert action in $U_4$}
\bea S &=& - \int d^4x \sqrt{-g}\; e^{2\Theta}R \n\\
&=& - \int d^4x \sqrt{-g}\; e^{2\Theta} \left(R^{(V_4)} + \dfrac{16}{3}\; \partial_\rho\Theta\; \partial^\rho \Theta - \tilde K_{\nu\rho\a} \tilde K^{\a\nu\rho}\right) \n\\&&+ \textrm{ surface term}\eea 
Here $\int d^4x e^{2\Theta} \sqrt{-g} D_\mu Q^\mu = \int d^4x \partial_\mu (e^{2\Theta} \sqrt{-g} Q^\mu)$ is a surface term and doesn't contribute to the equation of motion since variation of fields is taken zero at the boundary.

\section{Vacuum equations}

{\bf A. Variation of action S w.r.t. tracefree contortion tensor $\tilde K_{\nu\rho\a}$}
\bea \delta_{\tilde K} S &=& \int d^4x \sqrt{-g}\; e^{2 \Theta} \delta _{\tilde K} ({\tilde K}_{\nu\rho\a} {\tilde K}^{\a\nu\rho})\n\\ 
\Rightarrow \dfrac{\delta S}{\delta {\tilde K}_{\b\mu\s} (y)}   &=& 2 \int d^4x \sqrt{-g}\; e^{2 \Theta} 
\left(  \dfrac{\delta {\tilde K}_{\nu\rho\a} (x)}{\delta {\tilde K}_{\b\mu\s} (y)}  \right) {\tilde K}^{\a\nu\rho}\n\\
&=& 2 \int d^4x \sqrt{-g}\; e^{2 \Theta} \left(  \delta ^\b _\nu {\delta ^\mu _{[\rho}} \delta ^\s _{\a ]} \delta ^4(x-y) \right) {\tilde K}^{\a\nu\rho}\n\\&=& \sqrt{-g}\; e^{2 \Theta} ({\tilde K}^{\s\b\mu} - {\tilde K}^{\mu\b\s}) = 0\\
\Rightarrow {\tilde K}^{\s\b\mu} &=&  {\tilde K}^{\mu\b\s}\eea
This shows that the tracefree contortion tensor is symmetric in 1st and 3rd indices. Also from eqn.(\ref{con1}), its antisymmetric in 2nd and 3rd indices. Any tensor which has such symmetry properties, has all its components vanishing as shown below
\bea &&{\tilde K}^{\a\b\s} = - {\tilde K}^{\a\s\b} = - {\tilde K}^{\b\s\a} = {\tilde K}^{\b\a\s} = {\tilde K}^{\s\a\b} = - {\tilde K}^{\s\b\a} = - {\tilde K}^{\a\b\s} \n\\ &&\Rightarrow{\tilde K}^{\a\b\s} = 0\label{vac1} \eea
With ${\tilde K}_{\mu\nu\a} = 0$ in eqn. (\ref{con1}),
\be  K_{\mu\nu\alpha} =  \dfrac{2}{3} (g_{\mu\nu} Q_\alpha - g_{\mu\alpha} Q_\nu )\label{con2}\ee

\sp\ni
{\bf B. Variation of action S w.r.t. scalar potential $\Theta$}
\bea \delta_{\Theta} S &=& - \int d^4x \; 2 \sqrt{-g}\; e^{2\Theta} \delta\Theta \left(R^{(V_4)} + \dfrac{16}{3}\; \partial_\rho\Theta\; \partial^\rho \Theta - \tilde K_{\nu\rho\a} \tilde K^{\a\nu\rho}\right)\n\\ &&- \int d^4x \; \sqrt{-g}\; e^{2\Theta} \left(\dfrac{32}{3}\right) \partial _\mu (\delta\Theta)\partial _\nu \Theta g^{\mu\nu} \eea
Here 2nd integral \bea &&= \dfrac{32}{3} \; \int d^4x \; \sqrt{-g}\; e^{2\Theta} \delta\Theta \left(2 \partial _\mu \Theta \partial ^\mu \Theta +  \left\{^\a _{\a\mu}\right\} \partial ^\mu\Theta + \partial _\mu\partial ^\mu\Theta \right )\eea
\bea &&\Rightarrow - \dfrac{1}{2} \dfrac{e^{-2\Theta}}{\sqrt{-g}} \left.\dfrac{\delta S}{\delta\Theta}\right| _{\tilde{K} = 0} = R^{(V_4)} - \dfrac{16}{3}\; \partial_\mu\Theta\; \partial^\mu \Theta - \dfrac{16}{3}\; \partial_\mu \partial^\mu \Theta - \dfrac{16}{3} \left\{^\a _{\a\mu}\right\} \partial^\mu \Theta \n\eea
\bea \textrm{With } - \dfrac{1}{2} \dfrac{e^{-2\Theta}}{\sqrt{-g}} \left.\dfrac{\delta S}{\delta\Theta}\right| _{\tilde{K} = 0} &=& 0\ ,\qquad \qquad\n\\ &\Rightarrow & R^{(V_4)} + \dfrac{16}{3}\; \partial_\mu\Theta\; \partial^\mu \Theta - \dfrac{16}{3}\; D_\mu D^\mu \Theta = 0 \label{vac2}\qquad\qquad\eea
So, the equation of motion for $\Theta$ is \[ R^{(V_4)} + \dfrac{16}{3} (\partial_\mu\Theta \partial^\mu \Theta -  \square\Theta) = 0\ .\] Here, $\square = D_\mu D^\mu$ is the d'Alambertian operator $U_4$.

\sp\ni
{\bf C. Variation of action S w.r.t. metric tensor $g^{\mu\nu}$}
\bea && \dfrac{\delta S}{\delta g^{\eta\kappa}(y)} = \n\\
&&- \int d^4x \;   e^{2\Theta} \left(-\dfrac{\sqrt{-g}}{2} g_{\omega\sigma}\dfrac{\delta g^{\omega\sigma}(x)}{\delta g^{\eta\kappa}(y)}\right) \left(R^{(V_4)} + \dfrac{16}{3}\; \partial_\rho\Theta\; \partial^\rho \Theta - \tilde K_{\nu\rho\a} \tilde K^{\a\nu\rho}\right)\n\\ && - \int d^4x \;   e^{2\Theta} \sqrt{-g}\; \dfrac{\delta g^{\mu\nu}(x)}{\delta g^{\eta\kappa}(y)} \left(R^{(V_4)}_{\mu\nu} + \dfrac{16}{3}\; \partial_\mu\Theta\; \partial _\nu \Theta\right)\n\\&& - \int d^4x \;   e^{2\Theta} \sqrt{-g} \left(\;g^{\mu\nu} \dfrac{\delta R^{(V_4)}_{\mu\nu}(x)}{\delta g^{\eta\kappa}(y)} + \dfrac{\delta (\tilde K_{\nu\rho\a} g^{\a\b} g^{\mu\nu} g^{\rho\s} \tilde K_{\b\mu\s})}{\delta g^{\eta\kappa}(y)} \right) \eea
Using $\dfrac{\delta g^{\omega\sigma}(x)}{\delta g^{\eta\kappa}(y)} = \delta^{\o}_{(\eta}\delta^\s_{\kappa)}\delta^4 (x-y)$,
\bea && \dfrac{\delta S}{\delta g^{\eta\kappa}(y)} = e^{2\Theta} \dfrac{\sqrt{-g}}{2} g_{\eta\kappa} \left(R^{(V_4)} + \dfrac{16}{3}\; \partial_\rho\Theta\; \partial^\rho \Theta - \tilde K_{\nu\rho\a} \tilde K^{\a\nu\rho}\right)\n\\ && - e^{2\Theta} \sqrt{-g}\;  \left(R^{(V_4)}_{\eta\kappa} + \dfrac{16}{3}\; \partial_\eta\Theta\; \partial _\kappa \Theta\right) - \int d^4x \;   e^{2\Theta} \sqrt{-g}\;g^{\mu\nu} \overset{term (1)}{ \dfrac{\delta R^{(V_4)}_{\mu\nu}(x)}{\delta g^{\eta\kappa}(y)}}\n\\ && + \int d^4x \;   e^{2\Theta} \sqrt{-g}\;\overset{term (2)}{\dfrac{\delta (\tilde K_{\nu\rho\a} g^{\a\b} g^{\mu\nu} g^{\rho\s} \tilde K_{\b\mu\s})}{\delta g^{\eta\kappa}(y)}} \eea
$term (2) = \int d^4x \;   e^{2\Theta} \sqrt{-g}\;\tilde K_{\nu\rho\a} \tilde K_{\b\mu\s} \dfrac{\delta ( g^{\a\b} g^{\mu\nu} g^{\rho\s})}{\delta g^{\eta\kappa}(y)}$ vanishes with $\tilde K_{\a\b\s} = 0$.
\bea &\Rightarrow & - \dfrac{e^{-2\Theta}(y)}{\sqrt{-g(y)}}\left.\dfrac{\delta S}{\delta g^{\eta\kappa}(y)} \right|_{\tilde K_{\a\b\g} = 0} = -\dfrac{1}{2} g_{\eta\kappa} \left(R^{(V_4)} + \dfrac{16}{3}\; \partial_\rho\Theta\; \partial^\rho \Theta \right) + R^{(V_4)}_{\eta\kappa}\n\\&& + \dfrac{16}{3}\; \partial_\eta\Theta\; \partial _\kappa \Theta + \dfrac{e^{- 2\Theta}}{\sqrt{-g}} \int d^4x \;   e^{2\Theta} \sqrt{-g}\; \left.\left(g^{\mu\nu}  \dfrac{\delta R^{(V_4)}_{\mu\nu}(x)}{\delta g^{\eta\kappa}(y)}\right)\right| _{\tilde K = 0}\label{met1} \eea
To solve the last term, recall $\delta R_{\mu\nu}^{(V_4)} = \nabla _{\a}^{(V_4)} \delta\left\{^\a _{\mu\nu}\right\} - \nabla _{\mu}^{(V_4)} \delta \left\{^\a _{\a\nu}\right\}$. But in $U_4$, with the covariantly constant density as $e^{2\Theta} \sqrt{-g}$, we need to consider 
\be D _{\a} \;\delta \left\{^\a _{\mu\nu}\right\} - D _{\mu}\; \delta \left\{^\a _{\a\nu}\right\}\ee for evaluating the integral.
However, note that $\delta R_{\mu\nu}^{(U_4)} \textrm{or } \delta R_{\mu\nu} = D_\a \delta{\Gamma}^\a_{\mu\nu} - D_\mu \delta{\Gamma}^\a_{\a\nu}$.
\bea &&D _{\a} \; \delta \left\{^\a _{\mu\nu}\right\} - D _{\mu}\; \delta \left\{^\a _{\a\nu}\right\} = \delta R_{\mu\nu}^{(V_4)} - K{_{\a\rho}}^\a \;\delta \left\{^\rho _{\mu\nu}\right\} + K{_{\a\mu}}^\rho \;\delta \left\{^\a _{\rho\nu}\right\} \n\\
&& + K{_{\a\nu}}^\rho \delta \left\{^\a _{\mu\rho}\right\} + K{_{\mu\rho}}^\a \delta \left\{^\rho _{\a\nu}\right\} - K{_{\mu\a}}^\rho \delta\left\{^\a _{\rho\nu}\right\} - K{_{\mu\nu}}^\rho \delta \left\{^\a _{\rho\a}\right\} \eea
\bea \Rightarrow\delta R_{\mu\nu}^{(V_4)} &=&  D _{\a} \;\delta\left\{^\a _{\mu\nu}\right\} - D _{\mu}\; \delta \left\{^\a _{\a\nu}\right\} + K{_{\a\rho}}^\a \;\delta \left\{^\rho _{\mu\nu}\right\} - K{_{\a\mu}}^\rho \;\delta \left\{^\a _{\rho\nu}\right\}\n\\
&& - K{_{\a\nu}}^\rho \delta \left\{^\a _{\mu\rho}\right\} + K{_{\mu\nu}}^\rho \delta \left\{^\a _{\rho\a}\right\}\eea
\bea \Rightarrow  g^{\mu\nu}\delta R_{\mu\nu}^{(V_4)} &=& D _{\a} \left(g^{\mu\nu}\delta \left\{^\a _{\mu\nu}\right\}\right) - D _{\mu} \left(g^{\mu\nu} \delta \left\{^\a _{\a\nu}\right\}\right) - g^{\mu\nu} 2\partial _\rho\Theta \delta\left\{^\rho _{\mu\nu}\right\}\n\\&& - 2 K{_\a}^{\nu\rho} \delta\left\{^\a _{\rho\nu}\right\}  + 2 \partial ^\rho \Theta \delta \left\{^\a _{\rho\a}\right\}\eea
\bea &&\Rightarrow \int d^4x \;   e^{2\Theta} \sqrt{-g}\; \left.g^{\mu\nu}  \dfrac{\delta R^{(V_4)}_{\mu\nu}}{\delta g^{\eta\kappa}(y)}\right| _{\tilde K = 0} =\n\\ && \int d^4x \;   e^{2\Theta} \sqrt{-g}\left [ D _{\a} \left(g^{\mu\nu}\dfrac{\delta \left\{^\a _{\mu\nu}\right\}}{\delta g^{\eta\kappa}}\right) - D _{\mu} \left(g^{\mu\nu} \dfrac{\delta \left\{^\a _{\a\nu}\right\}}{\delta g^{\eta\kappa}}\right)\right. - \overset {term (i)}{2 g^{\rho\nu} \partial _\a\Theta\;\dfrac{\delta \left\{^\a _{\rho\nu}\right\}}{\delta g^{\eta\kappa}}} \n\\&&\qquad\qquad \qquad\qquad - \overset {term (ii)}{2 K{_\a}^{\nu\rho}\dfrac{\delta \left\{^\a _{\rho\nu}\right\}}{\delta g^{\eta\kappa}} } +\left. \overset {term (iii)}{2 \partial ^\rho \Theta \dfrac{\delta \left\{^\a _{\rho\a}\right\} }{\delta g^{\eta\kappa}}}\right]\label{met2}\eea
First two are the surface terms, using Gauss's divergence law in $U_4$. Assuming variation of field to be zero at the boundary, the variation of surface terms vanishes. $term (ii)$ also vanishes since $K{_\a}^{\nu\rho}$ is antisymmetric in $\nu\leftrightarrow\rho$ while $\left\{^\a _{\rho\nu}\right\}$ is symmetric in the two indices. Using
\bea \dfrac{\delta \left\{^\a _{\rho\nu}\right\}}{\delta g^{\eta\kappa}} &=& \dfrac{1}{2} \dfrac{\delta g^{\a\b}}{\delta g^{\eta\kappa}} (\partial _\rho g_{\b\nu} + \partial _\nu g_{\b\rho} - \partial _\b g_{\rho\nu}) \n\\ &&+\; \dfrac{1}{2}\; g^{\a\b} \left(\partial _\rho \dfrac{\delta g_{\b\nu}}{\delta g^{\eta\kappa}} +\partial _\nu \dfrac{\delta g_{\b\rho}}{\delta g^{\eta\kappa}} - \partial _\b \dfrac{\delta g_{\rho\nu}}{\delta g^{\eta\kappa} }  \right)
\eea
and, as can be easily seen, $ \dfrac{\delta g_{\a\b}}{\delta g^{\eta\kappa}} = - g_{\a\o} g_{\b\lambda} \dfrac{\delta g^{\o\lambda}}{\delta g^{\eta\kappa}} $, we get after somewhat lengthy calculation $term (i)$ in (\ref{met2}) 
\bea &&- 2 \int d^4x \;   e^{2\Theta} \sqrt{-g}\; g^{\rho\nu} \partial _\a\Theta\;\dfrac{\delta \left\{^\a _{\rho\nu}\right\}}{\delta g^{\eta\kappa}}= e^{2\Theta} \sqrt{-g} \left(- 4 \;\partial _\eta\Theta\;\partial _\kappa\Theta + g_{\eta\kappa} \left\{^\s _{\s\a}\right\} \partial ^\a\Theta \right. \n\\ && - 2 \partial _\eta \partial _\kappa\Theta + 2 g_{\eta\kappa} \partial _\a\Theta\;\partial ^\a\Theta + g_{\eta\kappa}\partial _\a\partial ^\a\Theta + \left. 2 \partial _\a\Theta \left\{^\a _{\eta\kappa}\right\}\right) \label{met3}\eea
And $term (iii)$ in (\ref{met2}), 
  \bea  &&2 \int d^4x e^{2\Theta}\sqrt{-g}\;  \partial ^\rho \Theta \dfrac{\delta \left\{^\a _{\rho\a}\right\} }{\delta g^{\eta\kappa}} =  \n\\&& e^{2\Theta} \sqrt{-g} \left ( 2 g_{\eta\kappa} \partial _\a\Theta\;\partial ^\a\Theta + g_{\eta\kappa} \left\{^\s _{\s\a}\right\} \partial ^\a\Theta +  g_{\eta\kappa} \partial _\a\partial ^\a\Theta\right)\eea
  So, 
  \bea \dfrac{e^{-2\Theta}}{\sqrt{-g}}\int d^4x e^{2\Theta} \sqrt{-g} \left. g^{\mu\nu}  \dfrac{\delta R^{(V_4)}_{\mu\nu}}{\delta g^{\eta\kappa}(y)}\right| _{\tilde K = 0} &=&   - 4 \;\partial _\eta\Theta\;\partial _\kappa\Theta + 2g_{\eta\kappa}\nabla _\a\partial ^\a\Theta  \n\\ && - 2 \nabla _\eta \partial _\kappa\Theta + 4 g_{\eta\kappa} \partial _\a\Theta\;\partial ^\a\Theta \label{met4} \eea
  Also, using eqn. (\ref{con2}), \be \left. D_\eta\partial _\kappa\Theta\right| _{\tilde K = 0} = - 2 \nabla _\eta \partial _\kappa\Theta + \dfrac{4}{3}\partial _\eta\Theta\;\partial _\kappa\Theta - \dfrac{4}{3}g_{\eta\kappa} \partial _\a\Theta\;\partial ^\a\Theta \ee
Finally, from eqns. (\ref{met1}) and (\ref{met4}), equation of motion for the $g_{\mu\nu}$ field is
\bea &&- \dfrac{e^{-2\Theta}}{\sqrt{-g}}\left.\dfrac{\delta S}{\delta g^{\eta\kappa}} \right|_{\tilde K_{\a\b\g} = 0} =\n\\&& R^{(V_4)}_{\eta\kappa} - 2 D _\eta \partial _\kappa\Theta -\dfrac{1}{2} g_{\eta\kappa} \left(R^{(V_4)} + \dfrac{8}{3}\; \partial_\rho\Theta\; \partial^\rho \Theta - 4 \square\Theta\right) = 0\label{vac3} \eea
Eqns. (\ref{vac1}), (\ref{vac2}) and (\ref{vac3}) are the $U_4$ vacuum equations. Taking trace of (\ref{vac3}), we get
\be  R^{(V_4)} + \dfrac{16}{3} \partial _\rho\Theta\; \partial^\rho \Theta =  6\square \Theta \ee 
Comparing it with (\ref{vac2}), \be R^{(V_4)} + \dfrac{16}{3} \partial _\rho\Theta\; \partial^\rho \Theta =  \square \Theta = 0\ee  
Other form of $U_4$ gravity equations for the vacuum 
\bea \tilde K_{\a\b\s} = 0 \n\\
\square\Theta = D_\mu \partial ^\mu \Theta = D_{\mu} Q^\mu = 0 \n\\
R^{(V_4)}_{\mu\nu} - 2 D _\mu Q _\nu +\dfrac{4}{3}\; g_{\mu\nu}Q_\rho Q^\rho  = 0 \label{vceq}
\eea
Since the equations of motion are of algebraic type, and not differential equations, torsion is clearly non-propagating. The traceless tensor $\tilde K _{\a\b\s} = 0$ and only the trace $Q_\mu$ can be non-vanishing in vacuum, outside matter distributions.

\sp\ni
Curvature and torsion are the surface densities of Lorentz transformations and translations, respectively \cite{trau}. Variation of Einstein Hilbert action of $U_4$ w.r.t. the metric gives 
\be R_{\mu\nu} - \dfrac{1}{2} g_{\mu\nu} R = T_{\mu\nu} . \ee 
$T_{\mu\nu}$ is the canonical stress energy tensor. Variation w.r.t. the torsion tensor $Q{_{\mu\nu}}^\rho$ gives 
\be Q{_{\mu\nu}}^\rho + \delta ^\rho _\mu Q {_{\nu \s}}^\s - \delta ^\rho _\nu Q {_{\mu \s}}^\s = k S{_{\mu\nu}}^\rho \label{spin}\ee
where $S{_{\mu\nu}}^\rho$ is the spin density tensor. In vacuum or outside matter, $S{_{\mu\nu}}^\rho = 0$ and hence $Q{_{\mu\nu}}^\rho = 0$ as is seen by contracting (\ref{spin}).
\bea &&g^\nu_\rho (Q{_{\mu\nu}}^\rho + \delta ^\rho _\mu Q {_{\nu \s}}^\s - \delta ^\rho _\nu Q {_{\mu \s}}^\s) = 0 \n\\&\Rightarrow & Q {_{\mu \s}}^\s = 0 \n\\ &\Rightarrow &  Q{_{\mu\nu}}^\rho = 0 \label{spinb}
\eea 
 Thus in ECT, torsion does not propagate in vacuum as seen from (\ref{vceq}). 
 Infact since the intrinsic spin is absent in vacuum, torsion is vanishing (\ref{spinb}) and hence (\ref{vceq}) reduce to the usual vacuum equations of general relativity, $Q_{\mu\nu\a} = 0$ and $R^{(V_4)}_{\mu\nu} = 0$. {\it In vacuum, both general relativity and Einstein Cartan Theory are identical}. In particular, Einstein Cartan Theory satisfies the equivalence principle in the vacuum. Advantage of Einstein Cartan Theory is that it allows a singularity free Universe model, while General relativity predicts that every model of the Universe must have a singularity in the past or in the future.

\section{Field equations in matter: Spinors in curved space}
Consider a classical field $\psi (x)$, representing matter sources in the flat Minkowski space $R_4$. Its Lagrangian density ${\cal L}_m = {\cal L}_m (\psi, \partial\psi, \eta)$ is assumed to depend upon the constant Minkowski metric $\eta_{\mu\nu}$, matter field and the gradient of the matter field. 
When the gravitational interaction is introduced, the matter Lagrangian has to be generalized to become a scalar under general coordinate transformations $x^\mu \rightarrow x'^\mu$. This can be achieved by minimal coupling procedure, i.e replacing the Minkowski metric with the world metric tensor $\eta_{\mu\nu}\rightarrow g_{\mu\nu}$  and the partial derivative with the covariant one, $\partial \rightarrow\nabla$. Also we must add to the matter Lagrangian, a  kinetic term for the gravitational field, ${\cal L}_g = R$ where R is the curvature scalar for $U_4$.

\sp\ni
The symmetry group of general relativity is the Lorentz group of local rotations and boosts. In special relativity, however, the group of symmetries is the global Poincar\'e group. Einstein Cartan theory, describing spinors in curved space, extends this symmetry group to local Poincar\'e transformations.

\sp\ni
{\bf Vielbein or Cartan formulation of general relativity} 

\sp\ni
Spinors transform under the spinor representation of the Lorentz group as \be \psi_a \rightarrow (\Lambda_{\frac{1}{2}})^b_a \;\psi_b \textrm{ in flat space, }\quad \Lambda_{\frac{1}{2}} = exp\left(-\frac{i}{2}\o_{a b} S^{a b}\right)\ee
$\Lambda_{\frac{1}{2}}$ is the finite spinor transformation matrix with $\o_{a b}$ as the parameter and $S^{a b}$ the generator (Here Latin indices $a, b,..$ denote flat space and Greek indices $\a, \b, ..$ denote curved). But their transformation rules are difficult to generalise to curved backgrounds. To couple gravity to spinors, its necessary to use vielbein formulation of general relativity.  In this, one considers a set of locally inertial coordinates  where the Lorentz behaviour of spinors can be applied and then translated back to the world (curved) coordinates. The frame field vielbein \be e^a_\mu (x_0) = \dfrac{\partial y^a (x_0;x)}{\partial x^\mu (x_0;x)}\ee transforms the Lorentz coordinates $y^a$ to curved $x^\mu$. Thus, vielbein connects tensor components $T_{a b c..}$ in local Lorentz frames (labeled using Latin indices) with tensor components $T_{\mu \nu\a..}$ in the spacetime frame (labeled using Greek indices). Viel stands for many, vielbein covers all dimensions.  For four dimensions, these frame fields are called tetrads or vierbeins. Inverse vielbein $e^\mu_a$ transforms from curved coordinates to the flat. Here, graviton is represented by the vielbein field instead of the metric. Following  orthonormality relations are satisfied by the vielbein field $e^a_\mu$
\bea &&g_{\mu\nu}(x) = e^a_\mu(x) e^b_\nu(x)\eta_{ab}\n\\
&&\eta^{ab} = e_\mu^a(x) e^b_\nu(x)g^{\mu\nu}(x)\label{viel1}\eea and by its inverse $e_a^\mu = g^{\mu\nu}\;\eta_{ab}\; e_\nu^b  $
\bea &&g^{\mu\nu} = e_a^\mu\; e_b^\nu\;\eta^{ab}\n\\
&&\eta_{ab} = e^\mu_a \; e_b^\nu \;g_{\mu\nu}
\eea 
\bea \textrm{Also }&&e^\mu_a\; e^b_\mu = \delta^b_a\n\\
&&e^\mu_a\; e^a_\nu = \delta^\mu_\nu\n\\
&&e = \sqrt{-g} = det(e^a_\mu)\eea
Thus vielbein is like square root of the metric. In $D$ dim, vielbien has $D^2$ independent components. But from eqn.(\ref{viel1}), the theory is invariant under a local Lorentz transformation acting on the vielbein $e'^a_\mu = e^c_\mu\; {\Lambda^a}_c$\; . To see that, we use ${\Lambda^a}_c\; {\Lambda^b}_d\; \eta_{ab} = \eta_{cd}$. Number of such independent transformations are $\dfrac{D(D-1)}{2}$ . Using up these gauge symmetries leaves us with the same number of independent components as the metric, i.e. $\dfrac{D(D+1)}{2}$ for the vielbein. 


\sp\ni
In general, given a world tensor $B_{\mu\nu}$, its corresponding components $B_{ab}$ in the flat tangent manifold can be  
obtained by directly contracting the indices with the vierbein 
fields $ B_{ab} = e^\mu_a e^\nu_b B_{\mu\nu} $ and vice versa. It is important to stress that if $B_\mu B_\nu$ is a world tensor, i.e. a tensor under general coordinate transformations, then $B_a B_c$ is a world scalar, but it transforms like a tensor with respect to the local Lorentz transformations. $B_\mu B^\mu$ is both a world scalar and a Lorentz scalar
\be B_\mu B^\mu = g^{\mu\nu} B_\mu B_\nu =  g^{\mu\nu} e_\mu^a e_\nu^c B_a B_c = \eta^{ac} B_a B_c = B_a B^a\ee

\sp\ni
In the absence of gravity, the world metric tensor reduces to the Minkowski metric, $g_{\mu\nu} = \eta_{\mu\nu}$, and the vierbein  field is given by $e^a_\mu= \delta^a_\mu$ and its inverse $e_a^\mu= \delta_a^\mu$

\sp\ni
Using the vierbein field, the Dirac matrices $\gamma^\mu(x)$ for 
the $U_4$ manifold can be defined as $\gamma^\mu = e^\mu_a \gamma^a$
where $\gamma^a$ are the (constant) flat-space Dirac matrices.

\sp\ni
The derivative of a geometrical object carrying the Lorentz indices, which are anholonomic indices, can be made covariant under local Lorentz rotations provided that a tangent space connection ${\o_\mu}^{ab}$ is introduced. ${\o_\mu}^{ab}$ is called the {\it spin connection} or the anholonomic connection. E.g. for a local Lorentz contravariant vector $A^b$ , which transforms as \[A'^b = {\Lambda^b}_c (x) A^c .\] Its partial derivative doesn't transform like a vector. Infact 
\bea (\partial_\mu A^b)' &=& \partial_\mu ({\Lambda^b}_c\; A^c)\n\\
&=&  (\partial_\mu {\Lambda^b}_c)\; A^c +  {\Lambda^b}_c\; \partial_\mu A^c  \eea
We can define however a Lorentz covariant derivative \be D_\mu A^b = \partial_\mu A^b + {\o_\mu}{^b}_c A^c\ee which transforms correctly as \be (D_\mu A^b)' = {\Lambda^b}_c\; D_\mu A^c \ee 
provided that the spin connection transforms inhomogenously as 
\be {\o'_\mu}^{ab} = {\Lambda^a}_c \;{\o_\mu}{^c}_k \;(\Lambda^{-1})^{k b} - {(\partial_\mu \Lambda)^a}_c (\Lambda^{-1})^{c b}\ee
\bea \textrm{So we have,}&&  (D_\mu A^b)' = \partial_\mu A'^b + {\o'_\mu}^{b d} A'_d \n\\ &&= \partial_\mu ({\Lambda^b}_c  A^c) + \{ {\Lambda^b}_c \;{\o_\mu}{^c}_k \;(\Lambda^{-1})^{k d} - {(\partial_\mu \Lambda)^b}_c (\Lambda^{-1})^{c d} \} ({\Lambda^m}_d  A_m) \n\\
&&=  (\partial_\mu {\Lambda^b}_c )A^c + {\Lambda^b}_c \partial_\mu A^c + {\Lambda^b}_c \;{\o_\mu}{^c}_k \; \eta^{k m} A_m - {(\partial_\mu \Lambda)^b}_c \;\eta^{c m} A_m \n\\&& = {\Lambda^b}_c\; D_\mu A^c\eea

\sp\ni 
Now, since $B_k A^k$ is a Lorentz scalar, imposing $ D_\mu(B_k A^k) =  \partial_\mu(B_k A^k)$ gives
\bea (D_\mu B_k) A^k + B_k (D_\mu A^k) &=& (\partial_\mu B_k) A^k + B_k (\partial_\mu A^k) \n\\ (D_\mu B_k) A^k + B_k  (\partial_\mu A^k + {\o_\mu}{^k}_c A^c) &=& (\partial_\mu B_k) A^k + B_k (\partial_\mu A^k) \n\\  (D_\mu B_k) A^k &=& (\partial_\mu B_k) A^k  -  B_c \; {\o_\mu}{^c}_k A^k
\eea
requires the covariant derivative of a Lorentz covariant vector to be\be  D_\mu B_k = \partial_\mu B_k - {\o_\mu}{^l}_k \; B_l \ee 
However, the total covariant derivative of a geometrical quantity carrying both flat and curvilinear indices is to be performed using both the anholonomic connection ${\o_\mu}^{a b}$ and holonomic connection $\Gamma^\a_{\mu\nu}$. The resulting derivative is then covariant 
under both local Lorentz and general coordinate transformations. 
Thus the covariant derivative of the vierbein field is 
\bea D_\mu e^a_\nu &=& \partial_\mu e^a_\nu + {\o_\mu}{^a}_c e^c_\nu - \Gamma^\a_{\mu\nu} e_\a^a \label{viel2} \eea
Note that $\o$ acts only on the flat indices while $\Gamma$ only on the curved ones. The expression (\ref{viel2}) transforms like a 2nd order  covariant tensor under a general coordinate transformation
\be D_\mu e^a_\nu \rightarrow \dfrac{\partial x^\a}{\partial x'^\mu}\; \dfrac{\partial x^\b}{\partial x'^\nu} \;D_\a e^a_\b \ee
and like a contravariant vector under a local Lorentz transformation
\be D_\mu e^a_\nu \rightarrow {\Lambda^a}_c \;D_\mu e^c_\nu \ee
In Einstein Cartan Theory, the vierbein field is assumed to be covariantly constant\be D_\mu e^a_\nu = 0 \ .\ee  
This is the constraint of zero torsion  with torsion $T^a = D e^a$  defined as the Yang-Mills curvature or field strength of the vierbein \cite{nastase}. It provides a relation between the two connections $\o$ and $\Gamma$. Moreover, from the metricity condition $D_\a g_{\mu\nu} = 0$, 
the spin connection is constrained to be antisymmetric in 
the last two indices
\bea D_\a g_{\mu\nu} &=&  D_\a (e^a_\mu\; e^b_\nu\;\eta_{ab})\n\\&=& e^a_\mu \;e^b_\nu\; D_\a \eta_{ab} \n\\ &=& e^a_\mu e^b_\nu (\partial_\a \eta_{ab} - {\o_\a}{^c}_a \eta_{cb}  - {\o_\a}{^c}_b \eta_{ac} ) = 0\n\\ &\Rightarrow&   \o_{\a b a}   + \o_{\a a b} =0 . \eea
Using this, since $D_\mu A^b = \partial_\mu A^b + {\o_\mu}^{b  c} A_c$ , hence    $D_\mu A_k = \partial_\mu A_k + {\o_\mu}{_k}^l\; A_l $ .

\sp\ni
In a Riemannian spacetime, the spin connection is not an independent field but rather is a function of the vierbein and its derivatives. However in the Riemann Cartan spacetime, the spin connection represents independent degrees of freedom associated with the non-zero torsion.

\sp\ni
Thus in presence of matter (fermions), the complete action for the Einstein Cartan theory is \be S = \int d^4x\; e \; e^{2\Theta} \left({\cal L}_m (\psi, D\psi, e) - \dfrac{R (e, \o)}{2\kappa}\right)\ee
where $\kappa = 8 \pi G$, $G$ being the gravitational constant, \be R (e,\o) = R^{(U_4)} = e^\mu_a \;e^\nu_b\; {R_{\mu\nu}}^{a b}(\o)\ .\ee 
Riemann tensor ${R_{\mu\nu}}^{a b}(\o)$ is the Yang-Mills curvature or field strength of the spin connection, $ R = d\o + \o \wedge \o $. 

\sp\ni
For the Dirac field coupled to gravity with torsion, the Lagrangian density is
\bea &&{\cal L}_m = e^\mu_a \; \overline{\psi} \gamma^a \left(\partial_\mu - \dfrac{i}{2}{\o_\mu}^{cd} \s_{c d}\right)\psi + e_{\mu a} K_{\a\b\rho}\;\epsilon ^{\mu\a\b\rho}\;\overline{\psi} \gamma^a \gamma_5 \psi\n\\ 
\textrm{with  }&& \quad\s^{c d} = \dfrac{i}{2} \gamma^{[c}\gamma^{d]}\ .\eea
In general, the energy momentum tensor is given by \be T_{\mu\nu} \equiv \dfrac{2}{\sqrt{-g}} \dfrac{\delta (\sqrt{-g} {\cal L}_m)}{\delta g^{\mu\nu}} \ee
and spin density tensor is \be S^{\mu\b\a} \equiv \dfrac{1}{\sqrt{-g}} \dfrac{\delta (\sqrt{-g} {\cal L}_m)}{\delta K_{\a\b\mu}} \ .\ee
{\it Gravitational field equations in presence of matter}
\bea \dfrac{1}{\sqrt{-g}} \dfrac{\delta (\sqrt{-g} R )}{\delta g^{\mu\nu}} &=& \kappa T_{\mu\nu}\label{mat1}\\
\dfrac{1}{2\sqrt{-g}} \dfrac{\delta (\sqrt{-g} R )}{\delta K_{\a\b\mu}} &=& \kappa S^{\mu\b\a}\label{mat2} \eea
Eqn. (\ref{mat1}) says that a matter-energy distribution curves the spacetime and eqn. (\ref{mat2}) says that a spin density distribution sources the torsion in spacetime. However {\it since the field equations relate torsion algebraically to the spin sources, as seen from eqn.(\ref{spin}),  torsion is non propagating in Einstein Cartan Theory}. Thus torsion is the source of a contact  interaction, i.e., a spinning particle cannot influence another spinning particle by means of torsion of the manifold. 
Torsion disappears immediately outside the spinning bodies. This is one of the main characteristics of the Einstein-Cartan theory and in this way torsion becomes physically interesting only at the microscopic level or macroscopically, when considering extremely collapsed matter. 

\sp\ni 
Nonetheless, if the gravitational Lagrangian is chosen in 
analogy to the standard gauge theory formalism then we are led to a Lagrangian quadratic in the curvature \be{\cal L} \propto {R_{\mu\nu}}^{a b}\;{R^{\mu\nu}}_{a b}\ .\ee It contains a kinetic term for the torsion and hence torsion becomes a propagating field. But a theory with such Lagrangian is different from Einstein Cartan theory and is no longer equivalent to general relativity even if the torsion is vanishing, i.e., in vacuum \cite{sabb}.

\section{Superstrings with intrinsic torsion} 
Superstring theory is a well studied candidate for quantum gravity theory. $D_p$-branes are intrinsic to the type II superstring theory, whose lowest energy state is  the type II supergravity. The bosonic sector of II A and II B consists of NS-NS and R-R states.  The NS-NS sector in both theories is the same, with the massless or lowest energy states consisting of spin 2 graviton $g_{MN}$ which is symmetric and traceless field, spin 1 Kalb-Ramond  $B_{MN}$ that is an   antisymmetric 2-form gauge field and spin 0 scalar field dilaton $\phi$ that is the trace part. Here $M, N$ indices run over $0,1,2,..,9$ since the critical dimension $D=10$. R-R sector for II A and II B is different as II A consists of $D_p$-branes with $p$ even while II B consists of those with $p$ odd. A stable $D_p$-brane carries R-R charge and couples electrically with the $C_{p+1}$ R-R gauge potential. Note that a $C_{p+1}$ R-R form field can always be traded off with its magnetic dual $C^{(M)}_{7-p}$ which couples magnetically with  $D_p$-brane. Thus in II A, we have $D_0$ coupling electrically with $C_1$, $D_2$ with $C_3$, while $D_4$ coupling magnetically with $C_3$, $D_6$ coupling magnetically with $C_1$ and a domain wall \footnote{Domain walls are branes with just one transverse direction} $D_8$ . In II B, we have $D_{-1}$ coupling electrically with $C_0$, $D_1$ with $C_2$, $D_3$ with $C_4$, while $D_5$ coupling magnetically with $C_2$, $D_7$ coupling magnetically with $C_0$ and a spacetime filling $D_9$ brane. These couplings to the R-R potentials are the well known Wess Zumino-type couplings. A natural electric coupling is given by 
\be I_{WZ} = \rho_{p} \int d^{p+1}x \, {\cal P}[C_{(p+1)}]\ , \ee 
where $\rho_{p}$ is the charge density of the brane and ${\cal P}[C_{(p+1)}]$ is the pullback of the (p+1)-form gauge potential on its worldvolume \cite{bachas}. A natural magnetic coupling is  given by \be \rho^{M}_{p} \int  \, {\cal P}[C^{M}_{(7-p)}]\  . \ee 
  Both the NS-NS and R-R closed strings propagate in the bulk of spacetime. The total action is a sum of the bulk or supergravity action, Dirac-Born-Infeld action and the Chern-Simon terms 
 \be
 S = S_{SUGRA} + S_{DBI} + S_{CS} .
 \ee 
 A constant Kalb-Ramond NS-NS $B$-field with components parallel to a $D$-brane can not be gauged away because whenever we vary $B_{MN}$ with a gauge parameter $\Lambda_M = (\Lambda_\mu , \Lambda_m)$, then we must simultaneously vary $A_\mu$ on the $D$-brane as follows \bea \delta B_{MN} &=& \partial_M \Lambda_N - \partial_N \Lambda_M \n\\ \delta A_\mu &=& - \Lambda_\mu \  .
 \eea 
 
 \ni Here Greek indices are used for coordinates along the brane and index $m$ for coordinates normal to it. Thus the fully gauge invariant combination is $F_{\mu\nu} + B_{\mu\nu} = \cal F_{\mu\nu}$. On the $D$-brane, $F_{\mu\nu}$ is not fully physical because it is not gauge invariant, but $\cal F_{\mu\nu}$ is the physical field strength \cite{zwie}.
 
 \sp\ni
 $D$-brane is defined as an object on which the fundamental open strings can end. A fundamental string carries electric charge or string charge for the Kalb-Ramond (KR) field, analogous to the electric charge carried by a particle for the Maxwell field. Mass dimension of KR field is $[B] = 2$, while the dimensionless KR field is given by ${\cal B}_{\mu\nu} =  2\pi\alpha' B_{\mu\nu}$.  Interaction or coupling of the  NS-NS closed string fields with $D_p$-brane is the same Dirac-Born-Infeld action as in bosonic string theory
 \be S_{D_p} = - T_p \int d^{p+1}x\ e^{-\phi}\sqrt{-det(g_{\mu\nu} + {\cal B}_{\mu\nu} + \bar{F}_{\mu\nu} )} \label{dbi} \ ,\ee 
 where $g_{\mu\nu}$ and $B_{\mu\nu}$ are the components parallel to the brane, $\bar{F}_{\mu\nu} = 2\pi\alpha' F_{\mu\nu}$  and $F_{\mu\nu}$ is the gauge field living on the brane. The coefficient $D_p$-brane tension is determined for $B=0$,  \be T_p (B = 0) = \dfrac{1}{g_s (2\pi)^p (\alpha')^{\frac{p+1}{2}}} \ .\ee 
Eq.\ref{dbi} is for slowly varying fields $f$ i.e., neglecting derivative terms $\sqrt{\kappa}\, \frac{\partial f}{f} << 1$. Here, $\kappa = 2\pi\alpha'$ is a parameter defining the size of a string \cite{seiberg}. The corresponding two-dimensional {\it non-linear sigma model action}  describes the propagation of strings in curved spacetime. The background field is  understood to be arising from condensation of infinite number of strings. Torsion is interpreted as the field  strength associated with the vacuum expectation value of the antisymmetric tensor field which appears in the supergravity multiplet \cite{bars}.           
   
 \sp\ni
 Kalb-Ramond field is viewed as an electromagnetic field on the $D$-brane with the spatio-temporal component $B_{0i}$ as the electric part and the $B_{ij}$ as the magnetic part. Let us consider a $D_p$-brane in presence of NS-NS torsion, while setting the dilaton field, all R-R fields and fermions to zero. There is a constant or global Kalb-Ramond component $B^z_{NS}$ as well as a dynamical or local Kalb-Ramond component $B^{nz}_{NS}$ pulled back to the brane $D_p$. The dynamical Kalb-Ramond component corresponds to superstring with intrinsic torsion. 
 The constant NS 2-form leads to an effective open string metric \cite{seiberg}, which serves as the background metric $G^{(NS)}_{\mu\nu}$ for the brane.

\sp\ni
In presence of the totally antisymmetric torsion on $D_p$-brane, the contortion tensor in eq. \ref{contort} becomes \bea {K_{\mu\nu}}^\alpha &=& - {H_{\mu\nu}}^\alpha = - G^{\a\l}_{(NS)} H_{\mu\nu\l} \n\\ K_{\mu\nu\a} &=& - H_{\mu\nu\a} \eea 
Since the trace of the totally antisymmetric torsion vanishes, so the Ricci scalar in eq. \ref{sca2} becomes   
\bea R^{(D_p)} &=& R^{(NS)} - K_{\rho\a\mu} K^{\a\mu\rho} \n\\ &=& R^{(NS)} - H_{\rho\a\mu} H^{\a\mu\rho} \label{ricci-brane} \eea  
So the $F$-string $-$ $D_p$-brane action in terms of closed string variables $g_{\mu\nu}$, $B_{\mu\nu}$, $g_s$ and commutative gauge field $A_\mu$ is  a sum of the DBI action and the bulk or supergravity action of dynamical KR field,  \be S = -  \dfrac{1}{g_s (2\pi)^p (\alpha')^{\frac{p+1}{2}}} \int d^{p+1}x\, \sqrt{-det(g_{\mu\nu} + {\cal B}_{\mu\nu} + \bar{F}_{\mu\nu} )} \ - \dfrac{1}{6\,C^2} \int d^{10}x\, H_{MNL} H^{MNL} \ .\label{F-Dp-cs}\ee
Here, $[C^2] = 6-p-1 = 5-p$. In terms of the open string variables $G_{\mu\nu}^{(NS)}$, $\theta_{\mu\nu} = (B^{-1})_{\mu\nu}$, $G_s$ and non-commutative gauge field $\hat{A}_\mu$, the $D_p$-brane action in presence of torsion comes out to be 
\bea S_{\hat{D}_p} = - \, \dfrac{1}{G_s (2\pi)^p (\alpha')^{\frac{p+1}{2}}}  \int d^{p+1}x\, \sqrt{-det(G_{\mu\nu}^{(NS)} + \kappa\hat{F}_{\mu\nu} )}\n\\
 + \dfrac{1}{6\,C^2} \int d^{p+1}x\, \sqrt{-G^{(NS)}}\,(R^{(NS)} - H_{\rho\a\mu} H^{\rho\a\mu})  \ , \label{F-Dp-os}\eea
 where we have used eq. \ref{ricci-brane} in the second term. First term in eq. \ref{F-Dp-os} is the open string analog of the DBI action \cite{seiberg}. 
 Seiberg-Witten  \cite{seiberg} showed that the ordinary (or commutative) Abelian gauge field $A$ with constant curvature $F$ and constant NS 2-form is equivalent to a noncommutative gauge field $\hat{A}$ with $\theta = \frac{1}{B}$. Thus the Born-Infeld part in eqns. \ref{F-Dp-cs} and  \ref{F-Dp-os} are equivalent. Further investigation of the deformation of $D_p$-brane in a weakly curved NS-NS background is studied by the author in \cite{rkosm} and a simple heuristic derivation of the open string metric in presence of torsion is suggested. 
 
 \section{Conclusion}
We have seen that Einstein Cartan theory is a theory of gravitation that differs minimally from the general relativity theory. In the ECT field equations, spin is algebraically related to torsion, so the torsion is non-propagating. It is seen from contracting the torsion equation that torsion tensor also vanishes if the spin tensor vanishes. So in vacuum or outside matter, torsion vanishes and the two theories are identical. However, in presence of  matter or fermion field, the spin sources non-propagating torsion. Effect of spin and torsion are significant only at very high densities of matter, but these densities are much smaller than the Planck density at which the quantum gravitational effects are believed to dominate. Possibly, Einstein Cartan theory will prove to be a better classical limit of a future quantum theory of gravitation than general relativity.

\sp\ni
We next realized a $F$-string $-$ $D_p$-brane set up by assuming a $D_p$-brane in presence of a dynamical background of Kalb-Ramond NS-NS field, while setting the dilaton field, R-R fields and fermions to zero. We use the formula that we obtained for Ricci scalar in $U_4$ manifold to determine the Ricci scalar on the $D$-brane in presence of the totally antisymmetric torsion. Thus we arrived at the $D_p$-brane action which describes a superstring with intrinsic torsion. 

\def\anp{Ann. of Phys.}
\def\cmp{Comm.Math.Phys. }
\def\prl{Phys.Rev.Lett. }
\def\prd#1{{Phys.Rev.} {\bf D#1}}
\def\jhep{JHEP\ {}}{}
\def\jaat{J.Astrophys.Aerosp.Technol.\ {}} {}
\def\cqg#1{{Class.\& Quant.Grav.}}
\def\plb#1{{Phys. Lett.} {\bf B#1}}
\def\npb#1{{Nucl. Phys.} {\bf B#1}}
\def\mpl#1{{Mod. Phys. Lett} {\bf A#1}}
\def\ijmpa#1{{Int.J.Mod.Phys.}{\bf A#1}}
\def\mpla#1{{Mod.Phys.Lett.}{\bf A#1}}
\def\rmp#1{{Rev. Mod. Phys.} {\bf 68#1}}
\def\ptep{Prog. of Theo.\& Exper.Phys.}
\def\jcap{J.Cosmo.\& Astropar.Phys.}

\end{document}